\documentclass[12pt]{article}
\usepackage{epsfig}
\usepackage{graphics}
\usepackage{amsmath, amsthm, amssymb}
\def\Re{{\cal R \mskip-4mu \lower.1ex \hbox{\it e}\,}}
\def\Im{{\cal I \mskip-5mu \lower.1ex \hbox{\it m}\,}}
\def\ie{{\it i.e.}}

\def\sub#1{_{\lower.25ex\hbox{$\scriptstyle#1$}}}
\def\tev{\,{\ifmmode\mathrm {TeV}\else TeV\fi}}
\def\gev{\,{\ifmmode\mathrm {GeV}\else GeV\fi}}
\def\mev{\,{\ifmmode\mathrm {MeV}\else MeV\fi}}
\def\mpl{\ifmmode M_{pl}\else $M_{pl}$\fi}
\def\mpl{\ifmmode \overline M_{Pl}\else $\bar M_{Pl}$\fi}
\def\to{\rightarrow}

\def\subw{_{\rm w}}
\def\mh{\ifmmode m\sbl H \else $m\sbl H$\fi}
\def\mch{\ifmmode m_{H^\pm} \else $m_{H^\pm}$\fi}
\def\mt{\ifmmode m_t\else $m_t$\fi}
\def\mc{\ifmmode m_c\else $m_c$\fi}
\def\mz{\ifmmode M_Z\else $M_Z$\fi}
\def\mw{\ifmmode M_W\else $M_W$\fi}
\def\mws{\ifmmode M_W^2 \else $M_W^2$\fi}
\def\mhs{\ifmmode m_H^2 \else $m_H^2$\fi}
\def\mzs{\ifmmode M_Z^2 \else $M_Z^2$\fi}
\def\mts{\ifmmode m_t^2 \else $m_t^2$\fi}
\def\mcs{\ifmmode m_c^2 \else $m_c^2$\fi}
\def\mchs{\ifmmode m_{H^\pm}^2 \else $m_{H^\pm}^2$\fi}
\def\ztwo{\ifmmode Z_2\else $Z_2$\fi}
\def\zone{\ifmmode Z_1\else $Z_1$\fi}
\def\mtwo{\ifmmode M_2\else $M_2$\fi}
\def\mone{\ifmmode M_1\else $M_1$\fi}
\def\tb{\ifmmode \tan\beta \else $\tan\beta$\fi}
\def\xw{\ifmmode x\subw\else $x\subw$\fi}
\def\ch{\ifmmode H^\pm \else $H^\pm$\fi}
\def\lum{\ifmmode {\cal L}\else ${\cal L}$\fi}
\def\inpb{\,{\ifmmode {\mathrm {pb}}^{-1}\else ${\mathrm {pb}}^{-1}$\fi}}
\def\infb{\,{\ifmmode {\mathrm {fb}}^{-1}\else ${\mathrm {fb}}^{-1}$\fi}}
\def\epem{\ifmmode e^+e^-\else $e^+e^-$\fi}
\def\ffbar{\ifmmode f\bar f \else $f\bar f$\fi}
\def\ccbar{\ifmmode c\bar c \else $c\bar c$\fi}
\def\bbbar{\ifmmode b\bar b \else $b\bar b$\fi}
\def\llbar{\ifmmode l\bar l \else $l\bar l$\fi}
\def\eebar{\ifmmode e\bar e \else $e\bar e$\fi}
\def\ppb{\ifmmode \bar pp\else $\bar pp$\fi}
\def\bsg{\ifmmode B\to X_s\gamma\else $B\to X_s\gamma$\fi}
\def\bsll{\ifmmode B\to X_s\ell^+\ell^-\else $B\to X_s\ell^+\ell^-$\fi}
\def\bstt{\ifmmode B\to X_s\tau^+\tau^-\else $B\to X_s\tau^+\tau^-$\fi}
\def\lamt{\ifmmode \tilde\lambda\else $\tilde\lambda$\fi}
\def\shat{\ifmmode \hat s\else $\hat s$\fi}
\def\that{\ifmmode \hat t\else $\hat t$\fi}
\def\uhat{\ifmmode \hat u\else $\hat u$\fi}

\newskip\zatskip \zatskip=0pt plus0pt minus0pt
\def\matth{\mathsurround=0pt}
\def\lsim{\mathrel{\mathpalette\atversim<}}
\def\gsim{\mathrel{\mathpalette\atversim>}}
\def\atversim#1#2{\lower0.7ex\vbox{\baselineskip\zatskip\lineskip\zatskip
  \lineskiplimit 0pt\ialign{$\matth#1\hfil##\hfil$\crcr#2\crcr\sim\crcr}}}

\def\grtsim{\,\,\rlap{\raise 3pt\hbox{$>$}}{\lower 3pt\hbox{$\sim$}}\,\,}
\def\lsim{\,\,\rlap{\raise 3pt\hbox{$<$}}{\lower 3pt\hbox{$\sim$}}\,\,}

\renewcommand{\thefootnote}{\fnsymbol{footnote}}

\begin{document}

\bibliographystyle{prsty} 
\thispagestyle{empty}
\renewcommand{\thefootnote}{\fnsymbol{footnote}}

\begin{flushright}
{\small
SLAC--PUB--11344\\
July 2005\\}
\end{flushright}

\vspace{.8cm}


\begin{center}
{\bf\large Determination of Littlest Higgs Model Parameters at the ILC\footnote{Work
supported by Department of Energy contract  DE-AC02-76SF00515.}
\footnote{e-mails: $^a$conley@stanford.edu,
$^b$hewett@slac.stanford.edu, and $^c$myphle@stanford.edu}}

\vspace{1cm}

John A. Conley$^a$, JoAnne Hewett$^b$, and My Phuong Le$^c$\\
Stanford Linear Accelerator Center, Stanford University,
Stanford, CA  94309\\

\medskip

\end{center}

\vfill

\begin{center}
{\bf\large
Abstract }
\end{center}

\begin{quote}
We examine the effects of the extended gauge sector of the Littlest Higgs
model in high energy \epem~collisions.  We find that the search reach in
$\epem \to \ffbar$ at a $\sqrt s = 500 \gev$ International Linear Collider
covers essentially the entire parameter region where the Littlest Higgs
model is relevant to the gauge hierarchy problem.  In addition, we show
that this channel provides an accurate determination of the fundamental
model parameters, to the precision of a few percent, provided that the LHC
measures the mass of the heavy neutral gauge field.  Additionally, we show
that the couplings of the extra gauge bosons to the light Higgs can be
observed from the process $\epem\to Z h$ for a significant region of
the parameter space.  This allows for confirmation of the structure of the
cancelation of the Higgs mass quadratic divergence and would verify the
little Higgs mechanism.
\end{quote}

\vfill

\begin{center}

\end{center}

\newpage



%
\pagestyle{plain}

\section{Introduction}
\label{intro}

The Standard Model (SM) of particle physics is a remarkably successful
theory.  It provides a complete description of physics at currently accessible
energies, and its predictions have been confirmed to high accuracy by all
high energy experiments to date.  An important piece of the SM remains unexplained--the
mechanism of electroweak symmetry breaking.  Precision measurements
and direct searches suggest that this mechanism involves a weakly coupled Higgs boson
with a mass in the range $114<m_H<208\gev$ at 95\% CL.  The Higgs mass parameter, however, is
quadratically sensitive to UV physics.  New physics at the \tev~scale is therefore
necessary to keep the Higgs light without fine-tuning.  This is known as the hierarchy
problem.  Three main classes
of models, supersymmetry, extra dimensions, and little Higgs,
have been proposed to address the hierarchy problem.  Which of these theories
Nature has chosen will be determined in the coming years as the Large Hadron
Collider and the International Linear Collider probe the \tev~scale.

The little Higgs models \cite{Arkani-Hamed:2002qy,
Arkani-Hamed:2001nc, Skiba:2003yf} feature the Higgs as a pseudo
Nambu-Goldstone boson of an approximate global symmetry which is
broken by a vev at a scale of a few~\tev.  The breaking is realized
in such a way that the Higgs mass only receives quantum corrections
at two loops.  In contrast to supersymmetry, the one-loop
contribution to the Higgs mass from a SM particle is canceled by a
contribution from a new particle of the {\it same} spin.  Little
Higgs theories thus predict the existence of new top-like quarks,
gauge bosons, and scalars near the \tev~scale.  The distinguishing
features of this model are the existence of these new particles and
their couplings to the light Higgs.  Measurement of these couplings
would verify the structure of the cancelation of the Higgs mass
quadratic divergences and prove the existence of the little Higgs
mechanism.

The most economical little Higgs model is the so-called ``Littlest
Higgs'' (LH) \cite{Arkani-Hamed:2002qy}, which we introduce here and
describe in more detail in Sec. \ref{sec1}. This scenario is based on
a non-linear sigma model with
an $SU(5)$ global symmetry, which is broken to the subgroup $SO(5)$
by a vev $f$.  The vev is generated by some strongly
coupled physics at a scale $\Lambda_S\sim 4\pi f$; possible UV completions
of little Higgs theories are discussed in \cite{Skiba:2003yf, Katz:2003sn}.
The $SU(5)$
contains a gauged subgroup $[SU(2)\times U(1)]^2$ which is broken by the vev
to the SM electroweak group $[SU(2)_L\times U(1)_Y]$. The global $SU(5)$ breaking
leaves 14 massless Goldstone bosons, four of which are eaten by the
gauge bosons of the broken gauge groups, giving these gauge bosons
a mass of order $f$. In particular, we have a heavy $Z$-like boson
$Z_H$ and a heavy photon-like boson $A_H$ which, as we will see, are
phenomenologically important.  The other ten Goldstone bosons make
up a complex doublet and a complex triplet which remain massless at
this stage.  Masses for the complex triplet are generated at the \tev-scale
by one-loop gauge interactions.  The neutral component of the complex doublet plays the role
of the SM Higgs.  Its mass term comes from a Coleman-Weinberg potential
and has quadratically divergent corrections only at two loops, giving
$\mu^2\sim f^2/16\pi^2$.  Thus the natural scale for $f$ is around a \tev.
If $f$ is much higher than a few \tev, the Higgs mass must again be finely
tuned and this model no longer addresses the hierarchy problem.

The phenomenological implications of little Higgs models have been
explored in \cite{Arkani-Hamed:2002qy, Hewett:2002px,Chen:2003fm, Huo:2003vd,
Csaki:2003si, Csaki:2002qg, Ros:2004ag, Han:2003wu}.  Constraints
arise from electroweak precision data as well as from indirect and direct
production at LEP-II and the Tevatron.
For example, in the Littlest Higgs scenario, the lack of discovery of the $A_H$, which is
expected to be quite light, puts a lower bound on
$f$ in the few~\tev~range.  Significant electroweak constraints
come from tree-level and loop deviations of the $\rho$-parameter
and the weak mixing angle $\sin^2 \theta_w$ from their SM values. Combining these gives
a limit $f\gsim 4\tev$ which is relatively parameter independent.
Many variants of little Higgs models exist in the literature which
lower this bound to $f\gsim1-2\tev$. 

In this paper we use the processes $\epem\rightarrow\ffbar$ and
$\epem\rightarrow Z h$ to investigate experimental limits from LEP
II data on the Littlest Higgs parameters, to evaluate the extent of
the International Linear Collider's search reach in LH parameter space, and to see
how accurately the ILC will be able to determine the LH parameters.
We will see that the ILC can substantially extend the discovery reach of the LHC.
In addition, we will also see that the bounds from $\epem\to\ffbar$ at LEP II
exclude a large part of the LHC's
search reach in the $pp\to Z_H\to Z_L h\to \ell^+\ell^-\bbbar$ channel. 
Complementary discussions of the Littlest Higgs model at the ILC and LHC
can be found in \cite{Ros:2004ag, Han:2003wu}.
In Sec.~\ref{sec1}, we discuss the Littlest Higgs model in
detail.  In Sec.~\ref{sec2}, we examine the process
$\epem\rightarrow\ffbar$ at LEP II and the ILC and determine how
accurately the ILC
will be able to measure the LH parameters.
In Sec.~\ref{sec3} we explore the LH parameter space using
the process $\epem\rightarrow Z h$ at the ILC.

\section{The Littlest Higgs model and its parameters}
\label{sec1}

In this paper, we are mainly concerned with the extended neutral gauge sector
present in the LH model.  While this scenario also includes a number
of parameters that arise from
the top and scalar sectors, in which there are a number of new
heavy particles, the observables of concern in our analysis only
depend on the three parameters present in the extended heavy gauge sector.
These are $f$, the vev or ``pion decay
constant'' of the nonlinear sigma model, which we discussed in the
Introduction, and two mixing angles.  Although we focus on the Littlest
Higgs model, we note that an enlarged gauge sector with rather generic
features is present in all little Higgs scenarios.

The vev $f$ characterizes the scale of the $SU(5)\rightarrow SO(5)$
breaking; the effective field theory of the 14 Goldstone bosons has
the Lagrangian
\begin{equation}
\lum_{\Sigma} = \frac{1}{2} \frac{f^2}{4} \mbox{Tr} \left|{\cal
D}_{\mu} \Sigma \right|^{2}, \label{siglag}
\end{equation}
where $\Sigma$ is a $5\times 5$ matrix parametrization of the
Goldstone boson degrees of freedom \cite{Arkani-Hamed:2002qy,
Han:2003wu}.  The covariant derivative contains the gauge bosons
associated with the gauged subgroup $[SU(2)\times U(1)]^2$,
$W_1$, $W_2$, $B_1$, and $B_2$;
\begin{equation}
{\cal D}_{\mu} \Sigma = \partial_{\mu} \Sigma - i \sum^2_{j=1}\left(
g_j(W_j \Sigma + \Sigma W_j^T) + g'_j (B_j \Sigma + \Sigma B^T_j)
\right).
\end{equation}

At the same time, the $[SU(2)\times U(1)]^2$ is also broken to $[SU(2)_L\times U(1)_Y]$,
and the gauge boson
mass eigenstates after the symmetry breaking are
\begin{equation}
\begin{split}
W=sW_1+cW_2,\;\;\;  &W'=-cW_1+sW_2, \\
B=s'B_1+c'B_2,\;\;\;  &B'=-c'B_1+s'B_2.
\end{split}
\end{equation}
The $W$ are the massless gauge bosons associated with the generators of $SU(2)_L$
and the $B$ is the massless gauge boson associated with the generator of $U(1)_Y$. The $W'$
and $B'$ are the massive gauge bosons associated with the four broken generators
of $[SU(2)\times U(1)]^2$, with their masses being given by
\begin{equation}
\label{heavygaugemasses}
m_{W'}=\frac{f}{2}\sqrt{g_1^2 + g_2^2}=\frac{g}{2sc}f,\;\;\;
m_{B'}=\frac{f}{2\sqrt{5}}\sqrt{{g_1'}^2 +
{g_2'}^2}=\frac{g}{2\sqrt{5}s'c'}f.
\end{equation}
The mixing angles
\begin{equation}
\label{angles}
s=\frac{g_2}{\sqrt{g_1^2 + g_2^2}}\;\;\mbox{and}\;\;
s'=\frac{g_2'}{\sqrt{g_1'^2 + g_2'^2}}
\end{equation}
relate the coupling strengths of the two copies of $[SU(2)\times U(1)]$.
These two angles together with $f$ are the three parameters of the model that are
relevant to our analysis. As we will see, the factor of $\sqrt{5}$
in the denominator of the expression for $m_{B'}$
will have important phenomenological consequences.

The Higgs sector contains a scalar triplet in addition to a SM-like
scalar doublet.  The doublet and triplet both obtain vevs.  The doublet
vev, $v$, brings about electroweak symmetry breaking (EWSB) as in the SM, and thus $v=246\gev$.
The triplet vev, $v'$, is related to $v$ by the couplings in the Coleman-Weinberg
potential.  Taking these to be $\mathcal{O}(1)$ gives the relation $v'\simeq v^2/2f$.

\begin{figure}
    \begin{center}
      \epsfig{file= 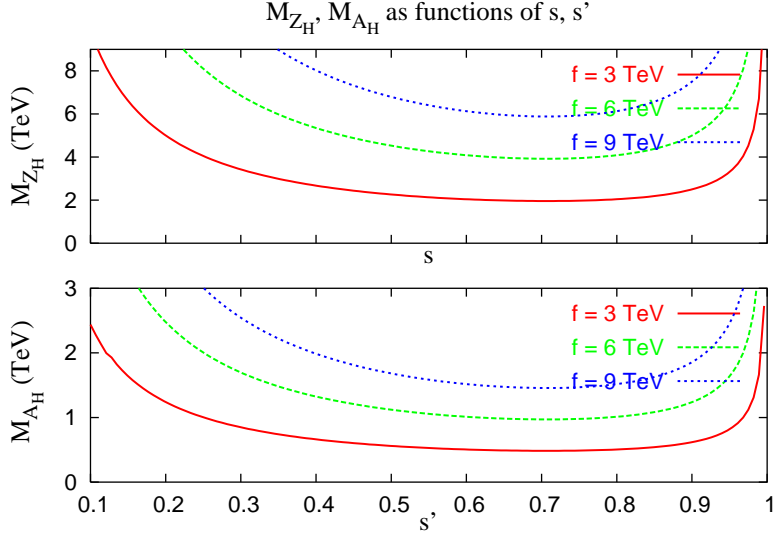}
      \caption{(Color online) Dependence of the heavy gauge boson masses $M_{Z_H}$ and $M_{A_H}$
      on $s$ and $s'$, respectively, for different values of $f$.}
      \label{mfig}
    \end{center}
 \end{figure}

After EWSB, the mass eigenstates
are obtained via mixing between the heavy ($W'$ and $B'$) and light ($W$ and
$B$) gauge bosons. They include the light (SM-like) bosons $W^{\pm}_L$,
$Z_L$, and $A_L$ observed in experiment, and new heavy bosons $W^{\pm}_H$, $Z_H$, and
$A_H$ that could be observed in future experiments.  At tree level, the processes $\epem\rightarrow\ffbar$ and
$\epem\rightarrow Z h$ involve the exchange of only the neutral gauge
bosons.  Their masses are given to $\mathcal{O}(v^2/f^2)$ by
\begin{equation}
\begin{split}
M^2_{A_L} &= 0,\\
M^2_{Z_L} &= m^2_Z\left[ 1 - \frac{v^2}{f^2}\left(\frac{1}{6}+
\frac{1}{4}(c^2-s^2)^2 + \frac{5}{4}(c'^2-s'^2)^2\right) +
8\frac{v'^2}{v^2}\right], \\
M^2_{A_H} &= m^2_Z s^2_w\left[\frac{f^2}{5s'^2c'^2v^2}-1+
\frac{v^2}{2 f^2}\left(\frac{5(c'^2-s'^2)^2}{2 s_w^2}-x_H\frac{g}{g'}
\frac{c'^2s^2 +c^2s'^2}{cc'ss'}\right)\right],\\
M^2_{Z_H} &= m^2_W\left[\frac{f^2}{s^2c^2v^2}-1+
\frac{v^2}{2 f^2}\left(\frac{(c^2-s^2)^2}{2 c_w^2}+x_H\frac{g'}{g}
\frac{c'^2s^2 +c^2s'^2}{cc'ss'}\right)\right],\\
\end{split}
\label{masses}
\end{equation}
where $m_W$ and $m_Z$ are the SM gauge boson masses, and $s_w$ ($c_w$)
represents the sine (cosine) of the weak mixing angle.
Here $x_H$, given by \cite{Han:2003wu}
\begin{equation}
x_H=\frac{5}{2}g g'
\frac{scs'c'(c^2{s'}^2+s^2{c'}^2)}{5g^2{s'}^2{c'}^2-{g'}^2s^2c^2},
\end{equation}
characterizes the mixing between $B'$ and $W'^3$ in
the $A_H$ and $Z_H$ eigenstates.
It is important to note that all but the first term in the square brackets
for $M^2_{A_H}$ and $M^2_{Z_H}$ are numerically insignificant.  
Thus $M^2_{A_H}$ depends strongly on $s'$ and not on $s$, and vice versa
for $M^2_{Z_H}$.  This dependence is shown in Fig. \ref{mfig}. Note
that the $A_H$ is significantly lighter than the $Z_H$ and can be as
light as a few hundred \gev; we will discuss the consequences of this below.

After EWSB, the couplings of the gauge bosons $Z_L$, $A_H$,
and $Z_H$ to fermions similarly depend on $s$, $s'$ and $f$ because
of the mixing between the fields.  If we demand that the U(1) be anomaly-free, which requires
$y_u=-2/5$ and $y_e=3/5$ in the notation of \cite{Han:2003wu},
the general structure of the couplings is
\begin{equation}
\label{couplings}
\begin{split}
g(A_L\ffbar)&=g_{SM}(A\ffbar),\\
g(Z_L\ffbar)&=g_{SM}(Z\ffbar)\left(1+\frac{v^2}{f^2}a_i(s,s')\right),\\
g(A_H\ffbar)&=b_i\frac{g'}{2s'c'}(\frac{1}{5}-\frac{1}{2}{c'}^2),\\
g(Z_H\ffbar)&=\pm\frac{gc}{4s},
\end{split}
\end{equation}
where $g_{SM}$ represents the relevant coupling in the SM.
A and Z are the SM photon and Z boson, and $a_i$ and $b_i$ are both $\mathcal{O}(1)$
where $i$ labels the species of fermion.

The existence of the heavy gauge boson-Higgs couplings is a hallmark of the
Littlest Higgs model.  They can be probed using the process $\epem\rightarrow Z_L H$
through the exchange of the $Z_L$, $Z_H$, and $A_H$.  The relevant couplings
are given by
\begin{equation}
\label{higgscoup}
\begin{split}
g(Z_{L\mu}Z_{L\nu}H)&=g_{SM}(Z_{\mu}Z_{\nu}H)\left(1+\frac{v^2}{f^2}a(s,s')\right),\\
g(Z_{L\mu}Z_{H\nu}H)&=\frac{-i}{2}\frac{g^2}{c_W}v\frac{c^2-s^2}{2sc}g_{\mu\nu},\\
g(Z_{L\mu}A_{H\nu}H)&=\frac{-i}{2}\frac{gg'}{c_W}v\frac{c'^2-s'^2}{2s'c'}g_{\mu\nu}.
\end{split}
\end{equation}
where $a$ is an $\mathcal{O}(1)$ function.  The formulae for the couplings can be found in
Appendix B of the first paper in \cite{Han:2003wu}.

Certain bounds on $s$ and $s'$ can be obtained by requiring that these
couplings remain perturbative.  Using the convention that a perturbative
coupling $g$ satisfies $g^2/4\pi<1$ gives
$s,s'\gsim0.1-0.2$.  Using the more conservative convention $g^2<1$ would give a smaller
allowed range for the parameters.
In the analysis that follows, we include the region where $s>0.16$.
As discussed above, expectations for the value of $f$ arise from the requirement of naturalness.
For $f\gsim 10\tev$, the LH model no longer addresses the hierarchy problem.

\begin{figure}
    \begin{center}
      \epsfig{file= 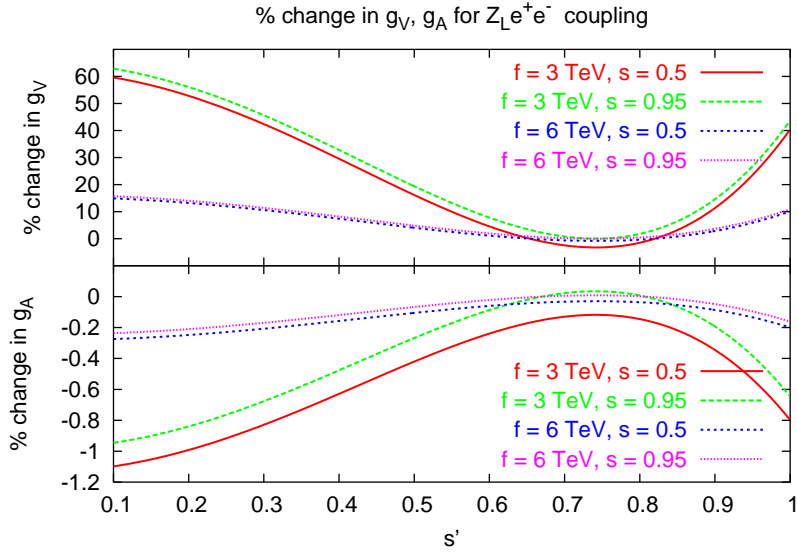}
      \caption{(Color online) The percent deviation
      of the vector and axial $Z_L \eebar$ couplings from the SM values for
      $Z_{SM}\eebar$, taking various values for the parameters f and s.}
      \label{gfig}
    \end{center}
 \end{figure}

As in \cite{Han:2003wu}, we write the fermion-boson coupling as
$i\gamma^{\mu}(g_V + g_A \gamma^5)$.  It turns out that for the electron-$Z_L$ coupling,
$|g_A| \gg |g_V|$, while in general the shifts in the couplings due to mixing are roughly equal,
{\it i.e} $|\Delta g_A| \simeq |\Delta g_V|$.  Thus the relative change in $g_V$ is in general much greater
than that for $g_A$, as shown in Fig. \ref{gfig}.
This relative change in $g_V$ is numerically fairly unimportant for
most of the observables in our analysis, as the cross sections are
typically functions of $g_V^2 + g_A^2$.  The left-right asymmetry
$A_{LR}$, however, has terms directly proportional to $g_V$.  Therefore,
for the ILC, which has beam polarization capability, the $A_{LR}$ deviation is important
and introduces a surprising $s'$ dependence in our results. We will discuss this in greater
detail in Sec. \ref{sec2}.

Equation~\ref{masses} shows that for generic choices of $s$ and $s'$,
$M_{A_H}/M_{Z_H}\simeq s_w m_Z / \sqrt{5} m_W \simeq 1/4$.  Figure~\ref{mfig}
illustrates this, with $M_{A_H}$ dipping well below 1~\tev~ for much of
the parameter space.  As mentioned in Sec.~\ref{intro}, this light $A_H$ is
responsible for the most stringent experimental constraints on the model
\cite{Hewett:2002px,Csaki:2002qg}.  As a result, phenomenologically viable
variations of the Littlest Higgs models typically decouple the $A_H$ by
modifying the gauge structure of the theory as in  \cite{Gregoire:2003kr}
and \cite{Kilic:2003mq}.  In this paper, however, we analyze the original Littlest Higgs
model as it is the most phenomenologically well-studied.  To gain some understanding of models in
which the $A_H$ decouples we take two approaches in our analysis.  One is to choose
a parameter value ($s'=\sqrt{3/5}$) for which the coupling of $A_H$ to fermions vanishes.
This decouples the $A_H$ from all tree-level electron-positron collider physics.  Another approach
is to artificially take $M_{A_H}\rightarrow\infty$ while letting all other
quantities in the theory take on their usual, parameter-dependent values.  While not
theoretically consistent, this approach gives us a more general picture of the behavior
of models in which the $A_H$ decouples.  We take both approaches and show the results
for each case throughout our analysis.

\section{Parameter Determination via \epem $\rightarrow$ \ffbar}
\label{sec2}

In this section we examine the process $\epem
\rightarrow \ffbar$, where all of the LH neutral gauge bosons
participate via s-channel exchange, at past and future colliders.
We first use a $\chi$-square analysis using the
$\epem \rightarrow \ffbar$ observables measured at LEP II.  This
analysis gives the region of LH parameter space excluded (to 95\%
confidence level) by the LEP II data.

We then perform a similar $\chi$-square analysis at the energies and
luminosity expected at the ILC.  We use the same set of observables
as in the LEP II analysis as well as the polarization asymmetries that will be measurable
due to the beam polarization capability at the ILC.  This analysis gives
the region of LH parameter space for which the ILC will be able to
determine (to 95\% confidence level) that the data cannot be
explained by the Standard Model, and represents the ILC
Littlest Higgs search reach.  The two analyses just mentioned are
described in Sec.~\ref{subsec21}.

Finally, in Sec.~\ref{subsec22}, we examine the ability of the ILC
to determine the values of the LH parameters from $\epem \rightarrow
\ffbar$.  For a few different generic sets of LH parameters, we
first generate sample data for the observables, and then perform a
$\chi$-square analysis to map out the region in LH parameter space
that is inconsistent (to 95\% CL) with the sample data.  The size
and shape of the remaining region tells us how accurately LH
parameters can be determined.

\subsection{The LEP II exclusion region and ILC search reach}
\label{subsec21}

Here we present our numerical analysis of the experimental
constraints on the Littlest Higgs parameter space from LEP II data
as well as the search reach expected from the ILC.  We use the
Lagrangian and Feynman rules of the Littlest Higgs model as
described in \cite{Han:2003wu}.  Note that for our analysis, we follow the
notation of \cite{Han:2003wu} and take
the values of the U(1) charge parameters $y_u = -2/5$ and $y_e =
3/5$ that, as previously discussed, leave the U(1) anomaly-free and
give the couplings shown in Eq.~\ref{couplings}. The remaining free parameters
of the model that are relevant to $\epem \rightarrow \ffbar$ are the
sines of the two  $[SU(2)\times U(1)]$ mixing angles, $s$ and $s'$;
and the ``decay constant,'' or vev, $f$ as defined in Eq.~\ref{siglag}
and Eq.~\ref{angles}.

We first study the constraints on the model from $\epem \rightarrow
\ffbar$ at LEP II, taking as our observables the normalized, binned
angular distribution and total cross section for $\epem
\rightarrow$\bbbar, \ccbar, and \llbar, with $l=e$, $\mu$, or $\tau$.  We use $\sqrt{s} = 200
\gev$ and an integrated luminosity \lum = 627
\inpb.  For the detection efficiencies, we
take $\epsilon_e$ = 97\%, $\epsilon_{\mu}$ = 88\%,
$\epsilon_{\tau}$ = 49\%,
$\epsilon_{b}$ = 40\%, and $\epsilon_{c}$ = 10\% ~\cite{Abbiendi:2003dh}.
We perform a $\chi$-square analysis and take the SM values for all
the observables to correspond to $\chi^2 = 0$, with a nonzero $\chi^2$
representing deviation from the SM. This is a reasonable approach,
since there was no
detectable deviation from the SM at LEP II
\cite{Abbiendi:2003dh}.

For the ILC analysis, in addition to the above mentioned observables,
we also include the angular binned left-right asymmetry $A_{LR}$
for each fermion pair. We use the projected energy $\sqrt{s} = 500\,\gev$
and luminosity \lum = 500 \infb. For the detection efficiencies, we
take $\epsilon_{e}$ = 97\%, $\epsilon_{\mu,\tau}$ = 95\%,
$\epsilon_{b}$ = 60\%, and $\epsilon_{c}$ = 35\%~\cite{Aguilar-Saavedra:2001rg}.

Because of the presence of the $Z_H$ and $A_H$, we use a general
formula for the differential cross section for $\epem \rightarrow
\ffbar$ that is valid for any set of extra gauge bosons that can run
in the s-channel \cite{Hewett:1988xc},
\begin{eqnarray}
\frac{d \sigma}{dz}=C_{\rm f} \frac{s}{32 \pi}
   \sum_{ij} P_{ij}^{ss}[ B_{ij}(1+z^2) + 2 C_{ij} z ],
\label{Form1}
\end{eqnarray}
where $z\equiv \cos\theta$, $C_{\rm f}$ is the color factor,
\begin{eqnarray}
 P_{ij}^{ss} \equiv \frac{(s-M_i^2)(s-M_j^2)+(\Gamma_i M_i)(\Gamma_j M_j)}
 {[(s-M_i^2)^2 +(\Gamma_i M_i)^2] [(s-M_j^2)^2 +(\Gamma_j M_j)^2]}, \label{Form2}
\end{eqnarray}
and
 $$ B_{ij}\equiv (v_i v_j+ a_i a_j)_f (v_i v_j+ a_i a_j)_e\, ,\;
 C_{ij}\equiv (v_i a_j+ a_i v_j)_f (v_i a_j+ a_i v_j)_e. $$
Here $v$ and $a$ correspond to the vector and axial couplings $g_V$
and $g_A$ discussed in Sec. \ref{sec1}, and the sum runs over the gauge
bosons in the s-channel: $A_L$, $Z_L$, $A_H$, and $Z_H$.

For Bhabha scattering, besides the s-channel, we also have a
contribution from the t-channel, so that
\begin{equation}
\begin{split}
\frac{d \sigma}{dz}= \frac{s}{32 \pi}
    \sum_{ij}  \bigg\{ & P_{ij}^{ss} \left[ B_{ij}(1+z^2) + 2 C_{ij} z \right] +\\
    & 2P_{ij}^{tt} \left[ B_{ij}\left(1 + \frac{1}{4}(1+z)^2\right)
   -C_{ij}\left(1 -\frac{1}{4}(1+z)^2\right)\right] +\\
    & P_{ij}^{st}(1+z)^2(B_{ij}+C_{ij}) \bigg\},
\label{Form3}
\end{split}
\end{equation}
where $P_{ij}^{st}$ and $P_{ij}^{tt}$ are defined similarly to
$P_{ij}^{ss}$ with the replacement $s\rightarrow t = -\frac{1}{2}s(1-z)$
in Eq.~\ref{Form2} in the obvious way.

To calculate $A_{LR}$, we need the cross sections for left- and right-handed electrons.
These can be obtained from the above formulae by making the replacements
\begin{equation}
v_{ie}\rightarrow\frac{1}{2}\left( v_{ie}+\lambda a_{ie}\right), \;
a_{ie}\rightarrow\frac{1}{2}\left( a_{ie}+\lambda v_{ie}\right),
\end{equation}
with $\lambda=+1(-1)$ corresponding to left- (right-) handed electrons.  Then the left-right
asymmetry is given by
\begin{equation}
A_{LR}(z)=\mathcal{P}\frac{\frac{d \sigma_L}{dz}-\frac{d \sigma_R}{dz}}{\frac{d \sigma_L}{dz}+\frac{d \sigma_R}{dz}},
\end{equation}
where $\mathcal{P}$ is the degree of $e^-$ beam polarization at the ILC, which we take to be 80\%.  We assume the $e^+$
beam is unpolarized.

We compute the $\chi^2$ distribution as follows, where $\sigma^i$ represents
one of the observables mentioned above:
\begin{eqnarray}
\chi^2=\sum_i \left(\frac{\sigma^i_{LH} - \sigma^i_{SM}}{\delta
\sigma^i}\right)^2.
 \label{Form4}
\end{eqnarray}
Here, $\delta\sigma$ is the statistical error for each observable, given by
\begin{equation}
\begin{split}
\delta\sigma_{tot}&=\sqrt{\frac{\sigma_{tot}}{\lum \epsilon}},\\
\delta(\frac{d\sigma_{N}}{dz})&=\sqrt{\frac{\frac{d \sigma_{N}}{dz}-(\frac{d \sigma_{N}}{dz})^2}{\lum \epsilon \sigma_{tot}}},\\
\delta A_{LR}&=\sqrt{\frac{1-A_{LR}^2}{\lum \epsilon \sigma_{tot}}},
\label{delta}
\end{split}
\end{equation}
where $\sigma_{tot}$ is the total cross section and $\frac{d\sigma_{N}}{dz}$ is the normalized differential cross section.
The efficiency $\epsilon$ for
each final state is given above.

As previously noted, $s,s',f$ are the free parameters present in the
neutral gauge sector of the
LH model. In our analysis, we choose a fixed value of $s'$ and scan the parameter space
$(s,f)$.

\begin{figure}
    \begin{center}
      \epsfig{file=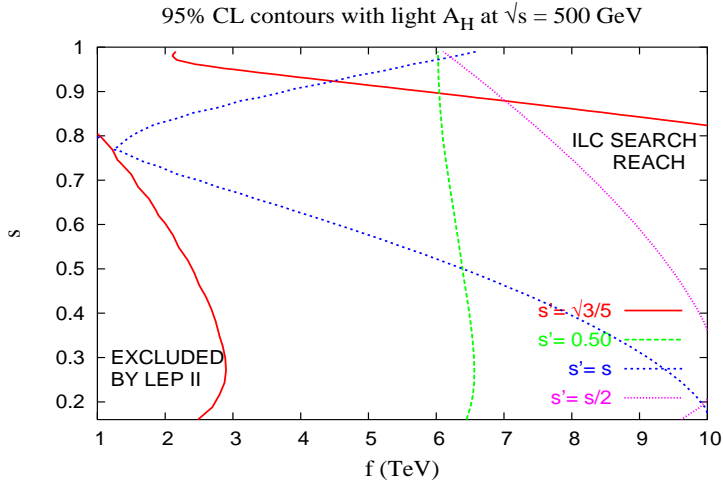,width=10.0cm,totalheight=6.5cm}
      \caption{(Color online) LEP II exclusion region from $\epem\rightarrow\ffbar$ and ILC search reach in the parameter space
    $(s,f)$ for different input values of $s'$, and including the $A_H$ contribution.
    For $s'=\sqrt{3/5}$ there are two
    lines of the same symbol/color, one on the boundary of the LEP II
    exclusion region, and one on the boundary of the ILC search reach
    region.  For the other values of $s'$ the curve shown is the boundary
    of the LEP II exclusion region, while the ILC search reach covers the entire
    parameter region shown.}
      \label{fig1}
    \end{center}
 \end{figure}

The exclusion region at LEP II and the search reach at the ILC correspond to
the regions where $\chi^2$ is greater than
5.99, representing a 95\% confidence level for two free
parameters.  The combined result is shown in
Fig.~\ref{fig1} for different values of $s'$.
For each value of $s'$, the LEP II exclusion region and the ILC search reach are on the left
of the corresponding contour.  This is because as $f$ increases, the
gauge boson masses (proportional to $f$) also increase (see Fig.~\ref{mfig}) and the
deviations in the $Z_L\ffbar$ couplings (proportional to $v^2/f^2$) decrease.
For the ILC search reach boundary one would expect to see four
contours at the upper right corner corresponding to the four different input
values of $s'$. However, there is only one
visible contour, for $s'= \sqrt{3/5}$, because in the other three
cases, the search reach covers the entire parameter space shown in
the figure.

As discussed above, the choice $s'= \sqrt{3/5}$ corresponds to decoupling the $A_H$
from the fermion sector, as verified by the
results shown in Eq.~\ref{couplings}.  In this case, the
ILC search reach can be as low as $f\sim 2 \tev$ for large values of
$s$. For other values of $s'$, the search reach is
greater than $f\sim 10 \tev$ for all values of $s$.  We thus see how
strongly the presence of the relatively-light $A_H$ can affect the
phenomenology. For LEP II, the story is similar; the exclusion
region for $s'= \sqrt{3/5}$ is much smaller than for other values of
$s'$.  This is because the observed deviation at $s'= \sqrt{3/5}$ is solely due to the
modification in the $Z_L\ffbar$ coupling and the presence of
the $Z_H$, which is generally several times heavier than the $A_H$.
For other values of $s'$ the constraints on $f$ can be as high as
$\sim$ 6 TeV. Overall, the LEP II
exclusion regions have constraints on the parameter $f$ that are
roughly consistent with those from precision measurements
\cite{Arkani-Hamed:2002qy, Hewett:2002px, Chen:2003fm,Huo:2003vd,
Csaki:2003si, Csaki:2002qg}.

\begin{figure}
    \begin{center}
      \epsfig{file=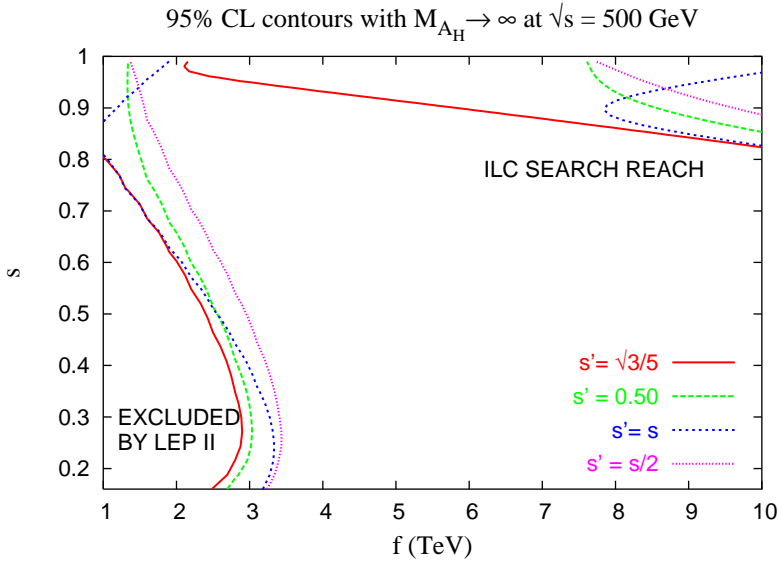}
      \caption{(Color online) LEP II exclusion region and ILC search reach as in Fig.~\ref{fig1}, but
      with $M_{A_H} \rightarrow \infty$ .}
      \label{fig2}
    \end{center}
\end{figure}

As discussed in Sec.~\ref{sec1}, we also examine the general behavior
of models in which the $A_H$ is decoupled by taking $M_{A_H}\rightarrow\infty$
while letting all other quantities take on their usual values.  The results
in this case are presented in Fig.~\ref{fig2}.
It is not surprising that the $s'=\sqrt{3/5}$ contours in
Fig.~\ref{fig2} are exactly the same as in Fig.~\ref{fig1}, since
the $A_H$ is decoupled for this choice of $s'$. For other values of $s'$, the
contours are very different in the two cases.
The s dependence of the contours in Fig.~\ref{fig2} is easy to understand.  The $Z_H
\ffbar$ couplings go as $gc/s$, thus the ILC contours show that the search reach is
higher for lower values of s. Similarly, for LEP II, the exclusion
region extends farther out in $f$ for lower values of s. There is,
however, a ``turnover'' for the LEP II exclusion region around
$s\sim 0.3$ where the contours start moving towards lower $f$.  This
takes place because the mass $M_{Z_H}$ begins to increase
(see Fig.~\ref{mfig}) and the overall contribution from $Z_H$ to
the observables starts to decrease as $s$ gets smaller.

\begin{figure}
    \begin{center}
      \epsfig{file=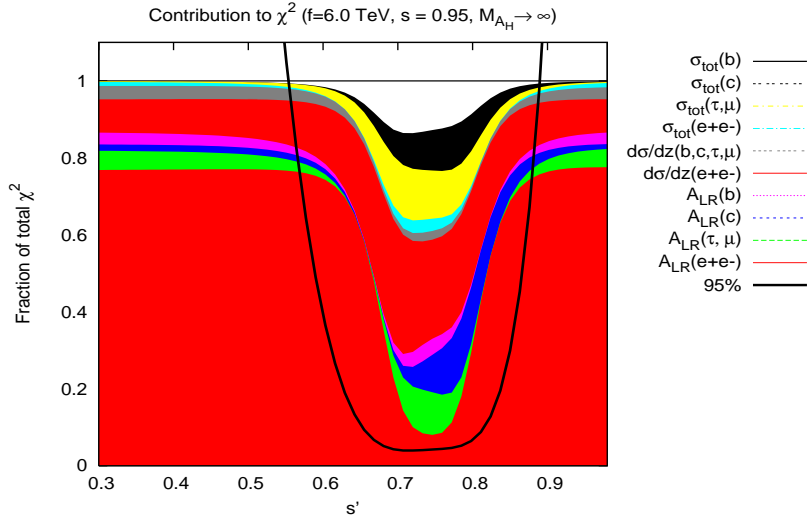, width=11.0cm,totalheight=7cm}
      \caption{(Color online) Fractional contribution to the total $\chi^2$ for
    each $\epem\rightarrow\ffbar$ observable at a $500 \gev$ ILC for fixed $(s,f)$.
    The labels on the legend go from top down in one-to-one
    correspondence with the shaded sections. For example, $A_{LR}(\epem)$
    contributes $\sim$ 78\% to the total $\chi^2$ at $s'=0.4$. The line
    labeled ``95\%'' is the total $\chi^2$/5.99.  This means that the region $s'\sim [0.55,0.9]$
    where this line dips below 1 is outside the $\sqrt{s}=500 \gev$ ILC search reach.}
      \label{fig3}
    \end{center}
 \end{figure}

With $M_{A_H}\rightarrow\infty$, the $s'$-dependence of the $\chi^2$
is mostly due to the deviation in the $Z_L\ffbar$ couplings, since neither
the $Z_H f\bar f$ couplings (see Eq.~\ref{couplings}) nor $M_{Z_H}$
(see Eq.~\ref{masses}) are strongly dependent on $s'$. This explains
why there is less variation among the different contours in Fig.~\ref{fig2}
than in Fig.~\ref{fig1}.  For values of $s$ close to $1$, however, the ILC contours
for different values of $s'$ begin to deviate from one another markedly.  This $s'$
dependence is due to the $s'$-sensitive deviation of $A_{LR}$, as discussed in Sec.~\ref{sec1}. This
is confirmed by Fig.~\ref{fig3}, which shows the relative contribution of the different observables to
the $\chi^2$ at the ILC with $M_{A_H}\rightarrow\infty$ and $s=0.95$.  Note that
$A_{LR}$ for various final states dominates the $\chi^2$ where it is large.

The fact that the
search reach is lowest for $s'=\sqrt{3/5}$ then indicates that the deviations
in the $Z_L\ffbar$ couplings are smallest for this parameter value.  This coincidence
arises because both $A_H$ and $Z_L$ are linear combinations of gauge eigenstates.  $A_H$
to lowest order is just $B'$, whose couplings to fermions vanish at $s'=\sqrt{3/5}$.  As the $s'$-dependent terms
in the deviation of the $Z_L\ffbar$ couplings arise from the $B'$ admixture, they also vanish at this value.
This is also confirmed by Fig. \ref{fig3},
which shows that the relative
contribution of $A_{LR}$ and the total $\chi^2$
decrease around $s'=\sqrt{3/5}$.

The search reach at a $\sqrt{s} = 1 \tev$ ILC reaches to somewhat higher
values of the parameter s, but has essentially the same reach for the parameter $f$ as
the $\sqrt{s} = 500 \gev$ machine.  This is reasonable; as $s$
approaches unity, the contribution from the deviations in the $Z_L$
couplings dominates the search reach, and these contributions are not as
important as the center of mass energy increases.  The result is that the
search reach is very simlar for both $\sqrt s = 500$ GeV and 1 TeV.  We
will see later, however, that the $\sqrt{s} = 1 \tev$ data can significantly
improve the parameter determination.

Figure~\ref{fig1} and Fig.~\ref{fig2} show that the $\sqrt{s}=500\gev$ 
ILC search reach in general covers most of the interesting parameter
space where the Littlest Higgs models are relevant to the gauge hierarchy.
Thus the $\epem \rightarrow \ffbar$ process
alone is an effective tool 
for investigating the Littlest Higgs model
at a  ILC.

\begin{figure}
    \begin{center}
      \epsfig{file=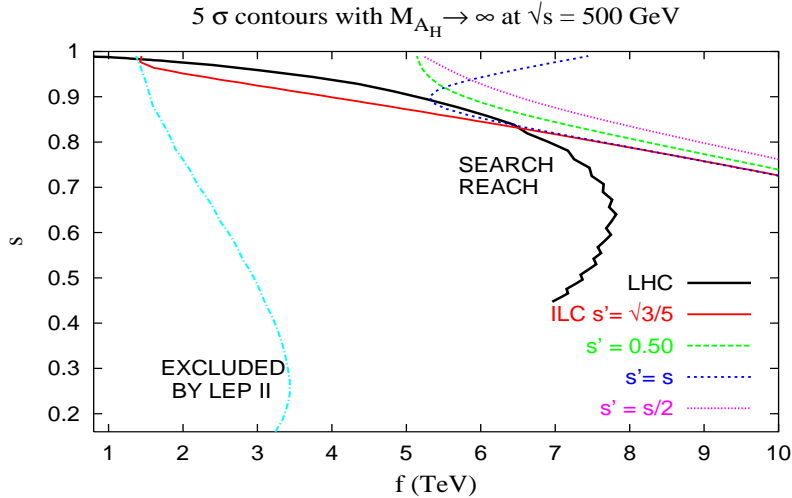, width=11.0cm,totalheight=7cm}
      \caption{(Color online) Comparison of ILC and LHC 
	search reach. The LHC data was taken from Fig. 8 of \cite{Azuelos:2004dm}.
        The search reach lies to the left of and underneath the contours.}
      \label{LHCfig}
    \end{center}
\end{figure}

It is important to compare the ILC search reach to that of the LHC.  An
ATLAS based analysis of the LHC search reach for the heavy gauge boson
$Z_H$ of the Littlest Higgs model was computed in 
Ref.~\cite{Azuelos:2004dm}.  The
resulting 5$\sigma$ contour for discovery of the $Z_H$ at the LHC is
reproduced in Fig.~\ref{LHCfig} 
(where we have converted their results to our choices
of parameters $f$ and $s$ for the axes). Fig.~\ref{LHCfig} also
displays our results for the ILC (taking $M_{A_H}\to\infty$),
where we have now employed a statistical
significance of 5$\sigma$ rather than 95\% to facilitate an equal
comparison.  We see that the ILC substantially extends the LHC search
reach for $s\lsim 0.8$.

\subsection{Parameter Determination: sample fits}
\label{subsec22}

We have now determined the available parameter space accessible to the
ILC and not already excluded by LEP II.  It remains to ask, given the
existence of an LH model with parameters in this accessible range, how
accurately would the ILC be able to measure them?  It is well-known
\cite{zprevs,ILCZp} that the ability to precisely measure the
couplings of heavy gauge bosons is one of the fortes of the ILC.

We first discuss the capability of the LHC to determine
the LH model parameters.  Numerous studies \cite{zprevs,LHCZp} have addressed
the ability of the LHC to determine the couplings of new gauge bosons.
The results of these studies show that while some model differentiation
is possible for $Z'$ bosons with masses $\lsim 2$~\tev, absolute
determination of the $Z'$ couplings is not possible.  There are three
main reasons for this: (i) the number of observables is limited in the
hadron collider environment.  The observables are the number of events
({\it i.e.}, cross section times branching fraction), the forward-backward
asymmetry, and the rapidity asymmetry) for leptonic final states only.
Other final states are not detectable above background, ($t\bar t$ final
states are a possible exception, but such events will be heavily smeared
and thus not useful for a coupling analysis).  (ii)  The observables
are convoluted with all contributing parton densities.  (iii)  The
statistics are insufficient for $M_{Z'}\gsim 2$ TeV.  Here, in the case
of the LH, our results show that LEP II essentially excludes the region
$M_{Z_H}\lsim 2$~\tev, and thus we do not expect the LHC to contribute
to the parameter extraction in any significant way.
We note, however, that a very precise mass measurement for $Z_H$ will be
obtained at the LHC.

To determine the accuracy of the parameter measurements, we perform some sample
fits, using a $\chi$-square analysis similar to the one described in
the preceding section.  We use the same ILC observables as before.  In some cases
we also include data from a $\sqrt{s}=1\tev$ run, for which we also take an 
integrated luminosity
of \lum = 500 \infb.  We note that we can exchange $M_{Z_H}$ for $f$, so
we now take $M_{Z_H}$, $s$, and $s'$ as our free
parameters.  We choose a generic data point $(s, s', M_{Z_H})$ and use it to
calculate the observables, which we then fluctuate according to
statistical error. We assume that the Large Hadron Collider would have determined
$M_{Z_H}$ relatively well, to the order of a few percent for $M_{Z_H}\lsim 5-6\tev$;
we thus fix $M_{Z_H}$ and perform a 2-variable
fit to $s$ and $s'$. Scanning the $s$-$s'$ parameter space, we
calculate the $\chi^2$ at every point. We find the minimum
$\chi^2$ point; the 95\% CL region surrounding it is the region for
which the $\chi^2$ is less than this minimum $\chi^2$ plus 5.99.

\begin{figure}
    \begin{center}
      \epsfig{file=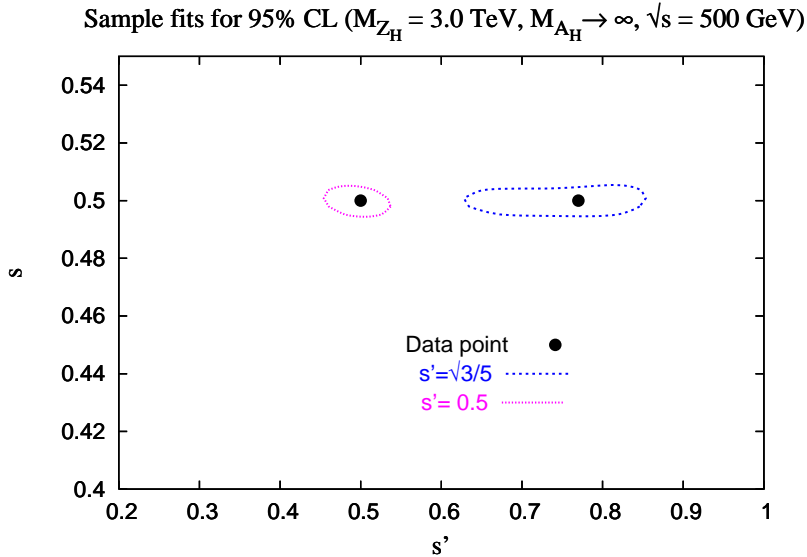}
      \caption{(Color online) 95\% CL sample fits to the data points ($s=0.5$, $s'=0.5$)
        and ($s=0.5$, $s'=\sqrt{3/5}$), using $\epem\rightarrow\ffbar$ observables
    at a 500 \gev~ILC, taking $M_{Z_H}=3.0\tev$ and
        $M_{A_H}\rightarrow\infty$.}
      \label{fit1}
    \end{center}
\end{figure}

Figure~\ref{fit1} shows the results of this fit for two sample data points
in the contrived scenario with $M_{A_H}\rightarrow\infty$.
For both of these points, the
determination of $s$ is very accurate.  This is due to the strong
dependence of the $Z_H \ffbar$ couplings on $s$, as discussed in the
previous section.  The $s'$ determination is worse than that for
$s$ because of our choice $s=0.5$.  
At this value of $s$, the contributions from
the $Z_L$ coupling deviations (which carry the $s'$ dependence) are
smaller than the $Z_H$ contributions.
The reason the
$s'$ determination is better for $s'=0.5$ than it is for
$s'=\sqrt{3/5}$ is that the $s'$-dependent deviations in $g_{V_{Z_L
\ffbar}}$ vanish for the latter value.

\begin{figure}
    \begin{center}
      \epsfig{file=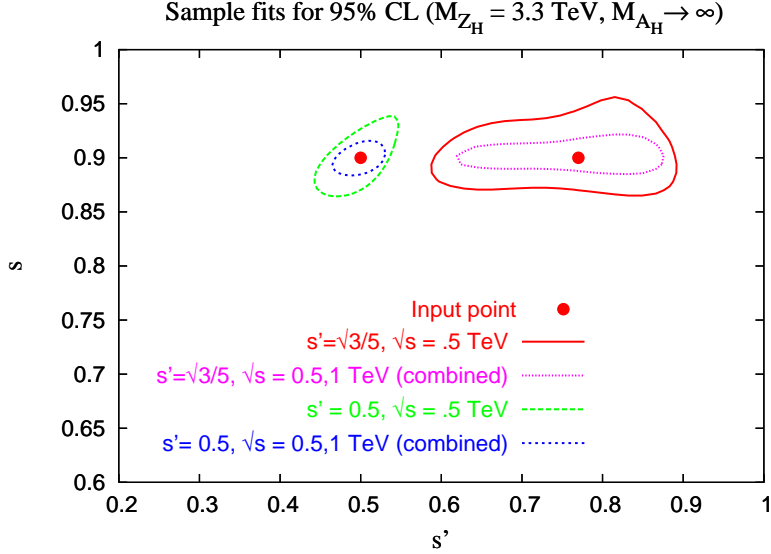}
      \caption{(Color online) Like Fig. \ref{fit1} except $M_{Z_H}=3.3\tev$ and the data points are ($s=0.9$, $s'=0.5$)
        and ($s=0.9$, $s'=\sqrt{3/5}$).  Also shown for each
        point is an improved fit from adding data from a $\sqrt{s}=1\tev$, \lum = 500 \infb run at the ILC.}
      \label{fit2}
    \end{center}
\end{figure}

Figure~\ref{fit2} shows the results from a similar fit and illustrates how it can
be improved with data from a higher-energy run with $\sqrt{s}=1\tev$ at the ILC.
Here, the $s$
determination is not much more accurate than the $s'$ determination, as the $s'$-independent
$Z_H$ contributions no longer dominate the fit for $s=0.9$.

\begin{figure}
    \begin{center}
      \epsfig{file=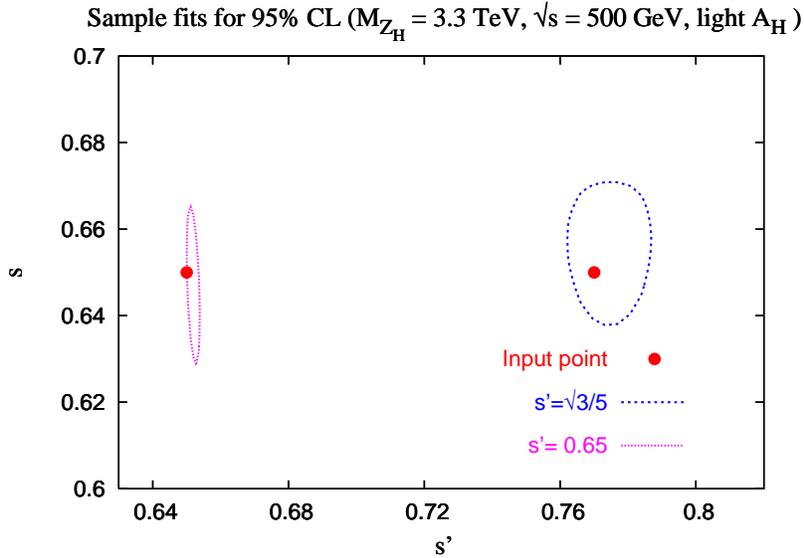}
      \caption{(Color online) Like Fig. \ref{fit1}, except $M_{Z_H}=3.3\tev$ and
	the data points are ($s=0.65$, $s'=0.65$)
        and ($s=0.65$, $s'=\sqrt{3/5}$), and
      the full $M_{A_H}$ contributions are included.}
      \label{fit4}
    \end{center}
\end{figure}

In Fig. \ref{fit4} we show results from a fit with the full $A_H$ contributions.
Not surprisingly, the parameter determination is much more precise, as the $A_H$
contributions, when present, dominate the $\chi^2$.  Since the $A_H$
couplings depend only on $s'$, it is also no surprise that here the $s'$ determination is
in general much better than that for $s$.

\begin{figure}
    \begin{center}
      \epsfig{file=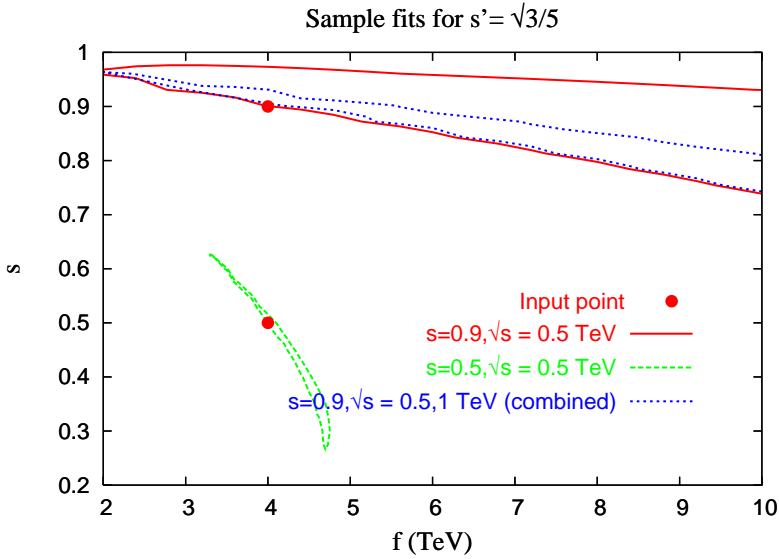}
      \caption{(Color online) Sample fits to the data points ($s=0.5$, $f=4\tev$)
        and ($s=0.9$, $f=4\tev$),  taking $s'=\sqrt{3/5}$.
        At the decoupling value of $s'$, the ${A_H}$ does not
        contribute.}
      \label{fit5}
    \end{center}
\end{figure}

If, for some reason, the LHC doesn't provide a good measurement of
$M_{Z_H}$, we would need to include that quantity, or equivalently
$f$, in our fits to the data.  In Fig. \ref{fit5} we show the results where we have
set $s'=\sqrt{3/5}$ and we fit to $s$
and $f$.  Note that for one of the data points,
the allowed region doesn't close.  This highlights the importance of using both the LHC
and the ILC to reliably determine the model parameters.

\section{Parameter determination using $\epem\rightarrow Z_L h$}
\label{sec3}

In order to confirm that the LH model is the correct description of
\tev-scale physics, it is important to test the hallmark of the LH
mechanism, namely that the Higgs mass quadratic divergences are
canceled by new particles with the same spin as their SM
counterparts.  The proof lies in the measurement of the new particle
couplings to the Higgs.  Here we are concerned with the coupling of
the heavy Z to the Higgs boson.  This coupling can be tested via the
process $\epem\rightarrow Z_L h$.  In the LH model, deviations of
observables related to this process from their SM expectations come
from three sources: the diagram with the $Z_H$ in the s-channel, the
diagram with the $A_H$ in the s-channel, and the deviation of the
$Z_LZ_Lh$ coupling from its SM value.

In this section we repeat the analysis of Section \ref{sec2}, using
the process $\epem\rightarrow Z_L h$ and taking the total cross section as our
observable with $m_h=120\gev$.  We assume that at a
$\sqrt{s}=500\gev$ ILC this cross section will be measured to an
accuracy of 1.5\% \cite{Aguilar-Saavedra:2001rg}.

The cross section, including the effects of additional gauge bosons, can be written as
\begin{eqnarray}
\sigma_{Z_L h}= \frac{|{\mathbf k}|}{8 \pi \sqrt{s}}
    \left(1+\frac{|{\mathbf k}|^2}{3 m_Z^2}\right)
   \sum_{ij} P_{ij}^{ss}[ {g_i}_{Z_L h} {g_j}_{Z_L h}(v_i v_j + a_i a_j)_e],
\label{zhform1}
\end{eqnarray}
where $P_{ij}^{ss}$ was defined in Eq.~\ref{Form2}.   The sum runs over the
participating gauge bosons in the s-channel: $Z_L$, $A_H$, and $Z_H$.  Here,
$|{\mathbf k}|$ is the magnitude of the 3-momentum of the
outgoing $Z_L$, given by
\begin{eqnarray}
|{\mathbf k}|=\frac{1}{2\sqrt{s}}\sqrt{(m_H^2-M_{Z_L}^2)^2 + s(s-2(M_{Z_L}^2+m_H^2))}.
 \label{zhform2}
\end{eqnarray}
The couplings $v_i$ and $a_i$ are the same as before--the axial and
vector couplings of electrons to the $i$th gauge boson.

\begin{figure}
    \begin{center}
      \epsfig{file=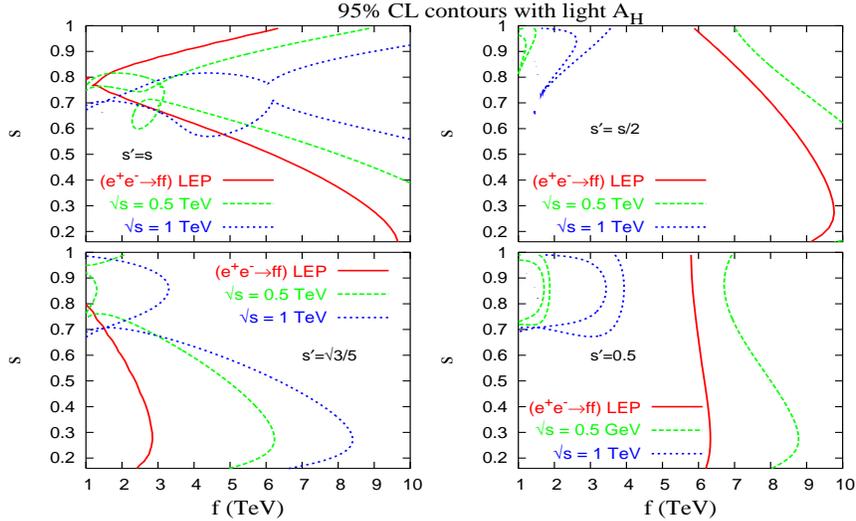,width=11.5cm,totalheight=7.0cm}
      \caption{(Color online) LEP II exclusion region from $\epem\rightarrow\ffbar$ and ILC search reach in the parameter space
    $(s,f)$ from the
    process $\epem\rightarrow Z_L h$, for various values of $s'$ and
    including the full $A_H$ contributions.  For each value of $s'$ there are three curves;
    one corresponds to the LEP II
    exclusion region, and the other two represent the ILC search reach
    region for $\sqrt{s}=500$ and $1000\gev$, respectively, taking an integrated
    luminosity of 500~\infb at each center-of-mass energy.}
      \label{Zh1}
    \end{center}
\end{figure}

We carry out the $\chi$-square analysis as before.
Figure~\ref{Zh1} shows the ILC search reach in the LH parameter space,
where each plot corresponds to a different choice
of $s'$.  By comparing
to Fig.~\ref{fig2} we note that $\epem\rightarrow Z_L h$ gives a much
poorer search reach than $\epem\rightarrow\ffbar$.  In particular, when $s'$
is near the decoupling value $\sqrt{3/5}$ the LH model is generally
indistinguishable from the SM.
Well away from $s'=\sqrt{3/5}$, as shown for $s'=s/2$ and
$s'=0.5$, the search reach covers almost all of parameter
space, except for regions of low $f$ where interference between the
$A_H$ and $Z_H$ conspire to bring the cross section near its SM
value.  These regions, however, are ruled out by LEP.

\begin{figure}
   \begin{center}
     \epsfig{file=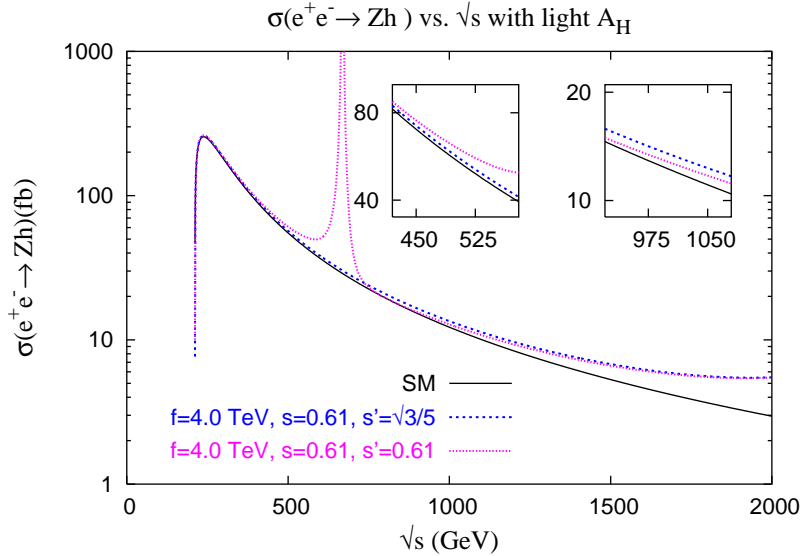}
     \caption{(Color online) The cross section for $\epem\rightarrow\ Z_L h$ as a function of $\sqrt{s}$ for
       the SM and two different points in LH parameter space.  The insets show the behavior near
       the expected ILC
       $\sqrt{s}$ values of $500\gev$ and $1\tev$.  The resonance at about 700~\gev~corresponds to the
       $A_H$.}
     \label{Zh4}
   \end{center}
\end{figure}

In the case $s'=s$, however, there are regions that exhibit similar
interference effects and are not ruled out by LEP data.  For
example, consider the two data points $f=4.0\tev$, $s=0.61$ with (a) $s'=\sqrt{3/5}$
and (b) $s'=0.61$.  With $\sqrt{s}=500\gev$, (b) is within the search reach while
(a) is just outside the search reach.  Figure~\ref{Zh4} shows that at this value of
$\sqrt{s}$, the deviation of the cross section from the SM is much greater for $s'=0.61$
than for $s'=\sqrt{3/5}$, since the $A_H$ decouples in the latter case.
With $\sqrt{s}=1\tev$, this behavior is reversed; point (a) is outside
the search reach while (b) is within.  At this value of $\sqrt{s}$ the interference between
$A_H$ and $Z_H$ brings the cross section closer
to the SM value when the $A_H$ contributes.

\begin{figure}
    \begin{center}
      \epsfig{file=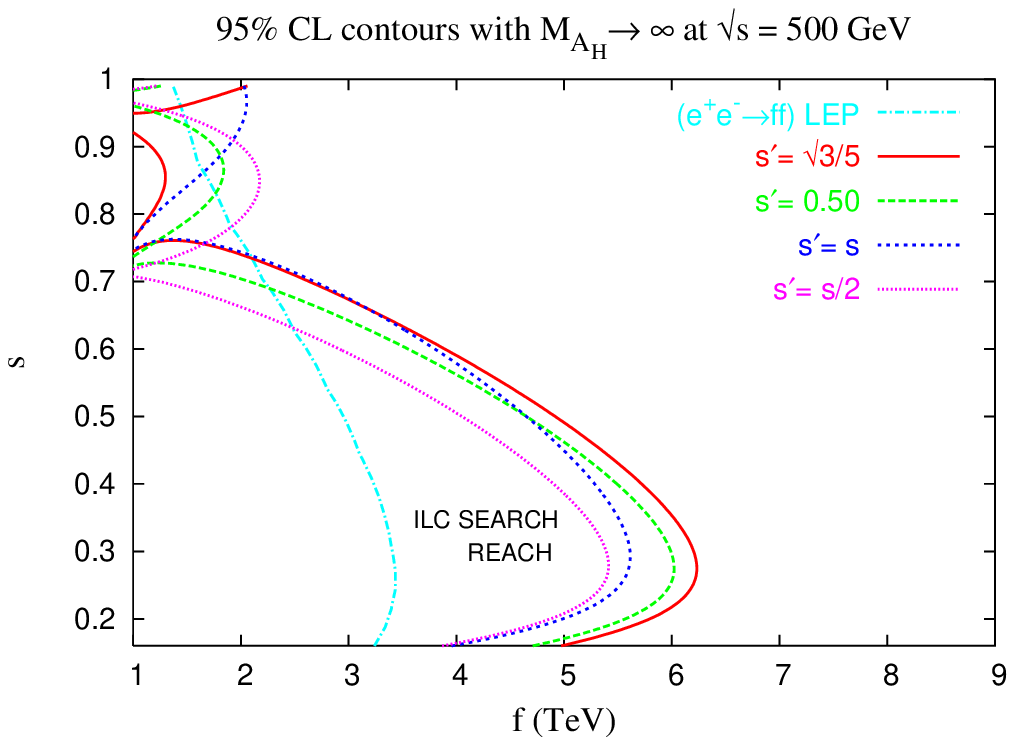}
      \caption{(Color online) The ILC search reach from the process $\epem\rightarrow Z_L h$
    for various values of $s'$, taking $\sqrt{s}=500$ and $M_{A_H}\rightarrow\infty$.
    The LEP II exclusion region from
    $\epem\rightarrow\ffbar$ is shown for $s'=s/2$
    (from Fig.~\ref{fig2}) for comparison.}
      \label{Zh2}
    \end{center}
\end{figure}

\begin{figure}
    \begin{center}
      \epsfig{file=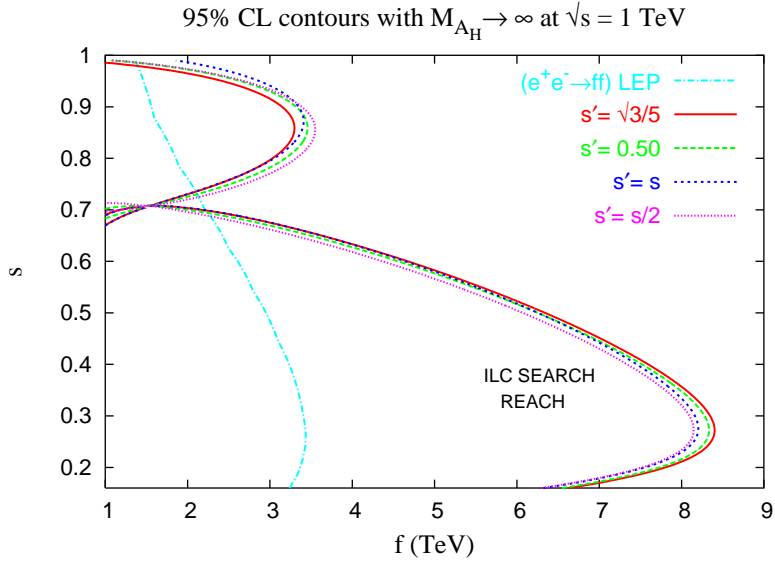}
      \caption{(Color online) Same as Fig.~\ref{Zh2}, but for $\sqrt{s}=1\tev$.}
      \label{Zh3}
    \end{center}
\end{figure}

Figure~\ref{Zh2} shows the search reach obtainable with 500~\infb
at a $\sqrt{s}=500\gev$ ILC,
taking $M_{A_H}\rightarrow\infty$.
Comparing to Fig.~\ref{fig1},
we see that the search reach here is much
smaller than for $\epem\rightarrow\ffbar$.  Figure~\ref{Zh3} displays the
corresponding reach at $\sqrt{s}=1\tev$ with 500~\infb.
In both cases, and for all choices of $s'$, the
search reach decreases markedly around $s=1/\sqrt{2}$.  This is
because the $Z_LZ_HH$ coupling vanishes at this value of $s$, as can be seen
in Eq.~\ref{higgscoup}.  It
is also interesting to note that the spread in the search reach as
$s'$ is varied is larger for $\sqrt{s}=500\gev$ than it is for
$1\tev$.  This can be understood if one notes that $\sqrt{s}=1\tev$ is closer
to the $Z_H$ pole (as $M_{Z_H}\simeq$ a few \tev ~throughout
the parameter space) than is $500\gev$.  Thus the deviation of
$\sigma_{Zh}$ from its SM value at $\sqrt{s}=1\tev$ is dominated by
the presence of the $Z_H$, whose mass and couplings are essentially
$s'$-independent.  At $\sqrt{s}=500\gev$, the deviation of
$\sigma_{Zh}$ has a more significant contribution from the deviation
of the $\epem Z_L$ coupling, which is strongly dependent on $s'$ (see
Fig.~\ref{gfig}).  For both values of $\sqrt{s}$, the sensitivity in the range of parameter space where
$s\gsim0.5$ does not reach beyond the general
precision electroweak bound of $f\gsim4\tev$.

One could hope to improve the sensitivity
by adding the measurement of the Higgs branching ratios as additional observables.  It turns
out, however, that the LH deviations of the branching ratios from their SM
values are at most 1-2\%, which is smaller than or equivalent to
the experimental sensitivity at the ILC.

\begin{figure}
    \begin{center}
      \epsfig{file=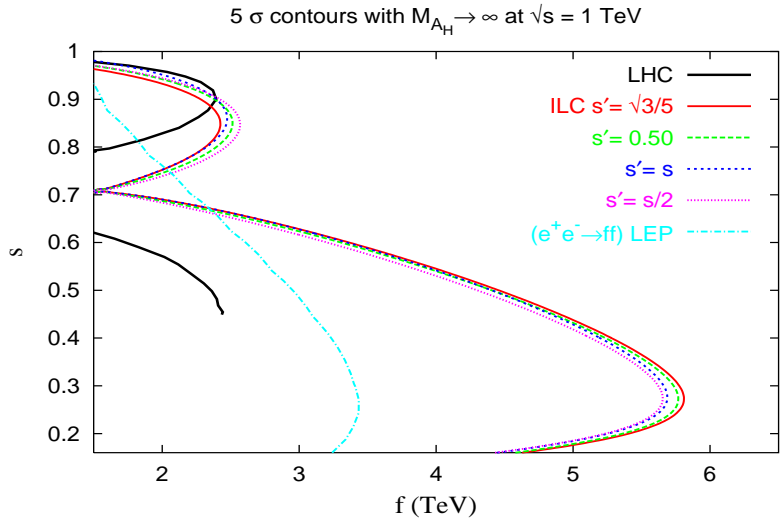, width=11.0cm,totalheight=7cm}
      \caption{(Color online) Comparison of ILC and LHC 
	search reach in the $Z_H\rightarrow Z h$ channel.  The LHC curve was read
        from Fig. 22 of \cite{Azuelos:2004dm}.}
      \label{LHCfig2}
    \end{center}
\end{figure}

Lastly, we again compare the reach obtainable at the ILC from this
process to that of the LHC in $pp\to Z_H Z_L+h$.  We display the
$5\sigma$ results
from the ATLAS based analysis \cite{Azuelos:2004dm} of this process in the LH using
the final state $\ell^+\ell^-b\bar b$ in Fig.~\ref{LHCfig2}.  We also show our
results, again adjusted for $5\sigma$ rather than $95\%$ statistical
significance.  This figure shows that the ILC overwhelms the
capability of the LHC in this channel.  In fact, our analysis of
$e^+e^-\to f\bar f$ shows that for $s\lsim 0.8$
the LEP II results already exclude the
possibility of the LHC observing the $Z_L+h$ decay of the $Z_H$.


%

\section{Summary}
Little Higgs models provide an interesting mechanism for addressing
the hierarchy problem.  They contain a single light Higgs boson
which is a pseudo-Goldstone boson with a small mass generated at the
two-loop level.  The quadratically divergent loop contributions to
the mass of this Higgs are canceled by contributions from new
particles appearing at the TeV scale.  These cancelations take place
between contributions from particles which have the same spin.
Measurement of the couplings of these new particles to the light
Higgs would verify the structure of these cancelations and establish
the Little Higgs mechanism.

Here, we have investigated the extended gauge boson sector within
these theories.  Numerous Little Higgs models, based on various global
symmetries, have been proposed.  However, the existence of an
enlarged gauge sector, with rather generic features, is endemic to all
these scenarios.  We choose to work
in the framework of the simplest model of this
type, known as the Littlest Higgs, based on an SU(5)/SO(5) nonlinear
sigma model.  This scenario contains the new heavy gauge bosons
$W^\pm_H$, $Z_H$, and $A_H$ in addition to the SM gauge fields.  The masses
of these additional gauge bosons are expected to be of order the
global symmetry breaking scale of $f\sim $ TeV.  (It is expected that
$f\lsim 10$ TeV in order for this scenario to be relevant to the hierarchy.)
However, due to the
group theory structure, the $A_H$ can be significantly lighter
resulting in stringent constraints from precision electroweak data.
Phenomenologically viable Littlest Higgs models must thus decouple
the $A_H$ and we have examined two such approaches in our analysis.
One, where we choose the model parameters such that the fermion couplings of
the $A_H$ vanish, and another where we artificially take
$M_{A_H}\to\infty$.

We study the effects of the new neutral gauge bosons in \epem\
annihilation.  These particles can participate in $\epem\to f\bar f$
and $\epem\to Z h$ via s-channel exchange, and their effects
can be felt indirectly for center of mass energies well below
their masses.   We find that fermion pair production is more sensitive
to Little Higgs effects than $Z h$ associated production.
We perform a thorough investigation of the model
parameter space and find that observables at LEP II exclude the region
$f\lsim 1-3$ TeV, which is consistent with the constraints obtained
from precision electroweak data.  The search reach of the proposed
International Linear Collider, operating at $\sqrt s=500$ GeV,
covers essentially the entire
parameter region where this model is relevant to the hierarchy,
\ie, $f\lsim 6-10$ TeV.  In the case of a 1 TeV ILC, the 
search region probes slightly larger values of the mixing
parameter $s$, but similar values of $f$.'

We have also demonstrated that once a signal is observed in these
channels, accurate measurements of the couplings of the heavy gauge
fields can be obtained from fermion pair production at the ILC.
These couplings are
related to the mixing angles in the extended gauge sector and we show
that experiments at the ILC can determine the fundamental parameters
of the theory.  For illustration, we performed a fit to generated
data for sample points in the Littlest Higgs parameter space, and
found that the fundamental parameters can be determined to the
precision of a few percent, provided that the LHC measures the
mass of the heavy neutral gauge field.  If information on the new
boson masses is not available from the LHC, then the parameter
determination at the ILC deteriorates.  Additionally, the couplings
of the extra gauge bosons to the light Higgs can separately be
determined from $\epem\to Z h$ for a significant region of
the parameter space.  This enables ILC experiments
to test the consistency of the theory and verify the structure 
of the Higgs quadratic divergence cancelations.

In summary, we find that the ILC has the capability to discover
the effects of the Littlest Higgs model over the entire theoretically
interesting range of parameters, and to additionally determine
the couplings of the heavy gauge bosons to the precision of a
few percent.

\subsection*{Acknowledgments}

We would like to thank T. Barklow, B. Lillie, H. Logan, M. Peskin, and
T. Rizzo for helpful discussions.  We thank G. Azuelos and G. Polesello for
providing data for Fig.~\ref{LHCfig}.



\end{document}